\documentclass[12pt]{article}
\usepackage{pic03}
\usepackage{hyperref}
\usepackage{url}
\usepackage{graphicx}
\usepackage{eurosym}

\def\ape{\textsf{APE}}
\def\anext{\textsf{apeNEXT}}
\def\amille{\textsf{APEmille}}

\begin{document}

\title{\bf 
 \anext{}: A MULTI-TFLOPS COMPUTER FOR SIMULATIONS IN LATTICE GAUGE THEORY
}

\author{%
  \renewcommand\thefootnote{\arabic{footnote}}
  \anext{}-Collaboration:
  F.~Bodin\footnotemark[\rennes],~
  Ph.~Boucaud\footnotemark[\orsay],~
  N.~Cabibbo\footnotemark[\roma1],~
  F.~Di~Carlo\footnotemark[\roma1],\\
  \renewcommand\thefootnote{\arabic{footnote}}
  R.~De Pietri\footnotemark[\parma],~
  F.~Di~Renzo\footnotemark[\parma],~
  H.~Kaldass\footnotemark[\ztn],~
  A.~Lonardo\footnotemark[\roma1],~
  M.~Lukyanov\footnotemark[\ztn],~
  S.~De~Luca\footnotemark[\roma1],\\
  \renewcommand\thefootnote{\arabic{footnote}}
  J.~Micheli\footnotemark[\orsay],~
  V.~Morenas\footnotemark[\clermont],~
  O.~Pene\footnotemark[\orsay],~
  D.~Pleiter\footnotemark[\nic],~
  N.~Paschedag\footnotemark[\ztn],~
  F.~Rapuano\footnotemark[\roma1],\\
  \renewcommand\thefootnote{\arabic{footnote}}
  D.~Rossetti\footnotemark[\roma1],
  L.~Sartori\footnotemark[\ferrara],~
  F.~Schifano\footnotemark[\ferrara],~
  H.~Simma\footnotemark[\ztn],~
  R.~Tripiccione\footnotemark[\ferrara],~
  P.~Vicini\footnotemark[\roma1]
}

\def\rennes{1}
\def\orsay{2}
\def\roma1{3}
\def\parma{4}
\def\ztn{5}
\def\clermont{6}
\def\nic{7}
\def\ferrara{8}

\footnotetext[\rennes]{IRISA/INRIA, Campus Universit\'{e} de Beaulieu, Rennes, France}
\footnotetext[\orsay]{LPT, University of Paris Sud, Orsay, France}
\footnotetext[\roma1]{INFN, Sezione di Roma, Italy}
\footnotetext[\parma]{Physics Department, University of Parma and
                             INFN, Gruppo Collegato di Parma, Italy}
\footnotetext[\ztn]{DESY Zeuthen, Germany}
\footnotetext[\clermont]{LPC, Universit\'{e} Blaise Pascal and IN2P3, Clermont, France}
\footnotetext[\nic]{NIC/DESY Zeuthen, Germany}
\footnotetext[\ferrara]{Physics Department, University of Ferrara, Italy}

\maketitle

\baselineskip=14.5pt
\begin{abstract}
We present the \ape{} (Array Processor Experiment) 
project for the development of dedicated parallel computers for numerical 
simulations in lattice gauge theories. While \amille{} is a production
machine in today's physics simulations at various sites in Europe, 
a new machine, \anext{}, is currently being developed to provide
multi-Tflops computing performance.
Like previous \ape{} machines, the new supercomputer is largely custom
designed and specifically optimized for simulations of Lattice QCD.
\end{abstract}

\baselineskip=17pt

\section{Introduction}

For many non-perturbative problems in quantum field theory, numerical simulations
on the lattice offer the only known way to compute various quantities from first 
principles.
Much progress has been made during recent years \cite{LAT}, e.g. in calculating 
the light hadron spectrum, the light quark masses, the running coupling
constant $\alpha_{\rm s}$ or observables in heavy quark physics.
Further-on, lattice simulations allow the study of phenomena, like chiral 
symmetry breaking and confinement, or of phase transitions in the early 
universe.

The computer resources required for such simulations are huge
and critically depend on the physical parameters, like quark 
masses, and the formulation of the theory on the lattice, e.g.~with 
improved chiral properties. To make the necessary
resources available, various research groups engage in the development
of massively parallel computers which are specifically optimized for
this kind of applications. One of these projects is APE \cite{APE} 
which is currently developing its fourth generation of machines, \anext{},
within the framework of an European collaboration by INFN (Italy),
DESY (Germany) and the University of Paris Sud (France).


\section{\anext{} Architecture}

\anext{} is designed as a massively parallel computer following
the Single Program Multiple Data (SPMD) programming model. 
All machine functionalities, including the network and memory 
interfaces, are integrated into one single custom chip running 
at a clock frequency of 200 MHz. The nodes run asynchronously
and are implicitly synchronized by communications. 

The processor core is a 64-bit architecture and all floating point
operations use IEEE double precision format. A so-called {\it normal}
operation $a\times b+c$, with $a$, $b$, and $c$ complex numbers, can
be started at each clock cycle. The peak performance of each node is 
therefore 1.6 GFlops.

Each node is a fully independent processor with 256--1024~MBytes 
private memory (standard DDR-SDRAM with ECC) storing both program and data.  
Since \anext{} has very long instruction words (VLIW), microcode 
de-compression and instruction buffers are used to reduce 
conflicts between data and instruction loading. A large number 
of 256 $(64+64)$-bit registers 
allows efficient data re-use.

The high-performance communication network has a 3-d torus topology. 
Each link moves one byte per clock cycle and the startup latency of about 
20 clock cycles, i.e.~100 ns, is very short. Communications and arithmetic 
operations
can be carried out in parallel. By loading local or remote data through 
dedicated pre-fetch queues one can almost perfectly exploit
the concurrency between memory accesses, network transfers, and arithmetic 
operations.

On an \anext{} processing board 16 nodes are mounted, and 
16 boards are interconnected on the backplane within a crate.
Larger systems are assembled by connecting together several crates using
external cables. Thus, a single rack system with 2 crates has 512 nodes 
and provides a peak performance of 0.8 TFlops. The footprint is about 1 m$^2$ 
only and the estimated power consumption is $< 10$~kW. Due to this very 
moderate power dissipation air cooling is still possible.

\anext{} systems are accessed from a front-end PC via a custom designed 
host-interface PCI-board. It uses a simple I2C link for bootstrapping 
and basic operating system requests, while I/O operations are handled 
via high-speed data links. By connecting one or more processing boards
of an \anext{} machine with host-interface boards the overall I/O 
bandwidth can be adjusted to the user needs.

Two high-level programming languages, TAO and C (with suitable
extensions for parallelisation) are supported on \anext{}. 
After further optimisation of the intermediate assembly, the microcode 
is generated, scheduled and compressed.


\section{Status and Outlook}

The design of \anext{} has been finished and prototype processors are to be
delivered in autumn 2003.  All other hardware components
exist and have been tested. 
The final procurement costs are estimated to be 0.5~\euro{}/MFlops 
peak performance.

The full VHDL design of the machine has been used for
benchmarking. Key lattice gauge theory kernels were measured to 
run at a sustained performance of $O(50\%)$ or more.
While the \anext{} design is optimized for simulating lattice QCD, we 
expect the architecture to be suitable also for other applications.

European research groups in lattice gauge theory aim for large dedicated 
installations of massively parallel computers during the next years. 
In Edinburgh (UK) a 10~TFlops QCDOC machine will be installed
by 2004 and in Italy a 10~TFlops \anext{} installation is planned.
In Germany researchers from various groups joined the
\textsf{Lattice Forum} (LATFOR) initiative. They have set up a broad
research program ~\cite{LatFor}, assessed the required computing 
resources, and proposed a 15 and 10~TFlops installation of 
\anext{} hosted by NIC at DESY (Zeuthen) and GSI (Darmstadt).


\end{document}